Hindawi

*Research Article*

# Publication Bias in Meta-Analysis: Confidence Intervals for Rosenthal's Fail-Safe Number

## Konstantinos C. Fragkos,[1,2] Michail Tsagris,[3] and Christos C. Frangos[4]

[1] *Department of Economics, Mathematics and Statistics, Birkbeck, University of London, Malet Street, London WC1E 7HX, UK*
[2] *Division of Medicine, University College London, Rockefeller Building, 21 University Street, London WC1E 6JJ, UK*
[3] *College of Engineering and Technology, American University of the Middle East, Egaila, Kuwait*
[4] *Department of Business Administration, Technological Educational Institute (T.E.I.) of Athens, 122 43 Athens, Greece*

Correspondence should be addressed to Konstantinos C. Fragkos; constantinos.frangos.09@ucl.ac.uk

The purpose of the present paper is to assess the efficacy of confidence intervals for Rosenthal's fail-safe number. Although Rosenthal's estimator is highly used by researchers, its statistical properties are largely unexplored. First of all, we developed statistical theory which allowed us to produce confidence intervals for Rosenthal's fail-safe number. This was produced by discerning whether the number of studies analysed in a meta-analysis is fixed or random. Each case produces different variance estimators. For a given number of studies and a given distribution, we provided five variance estimators. Confidence intervals are examined with a normal approximation and a nonparametric bootstrap. The accuracy of the different confidence interval estimates was then tested by methods of simulation under different distributional assumptions. The half normal distribution variance estimator has the best probability coverage. Finally, we provide a table of lower confidence intervals for Rosenthal's estimator.

## 1. Introduction

Meta-analysis refers to methods focused on contrasting and combining results from different studies, in the hope of identifying patterns among study results, sources of disagreement among those results, or other interesting relationships that may come to light in the context of multiple studies [1]. In its simplest form, this is normally done by identification of a common measure of effect size, of which a weighted average might be the output of a meta-analysis. The weighting might be related to sample sizes within the individual studies [2, 3]. More generally there are other differences between the studies that need to be allowed for, but the general aim of a meta-analysis is to more powerfully estimate the true effect size as opposed to a less precise effect size derived in a single study under a given single set of assumptions and conditions [4]. For reviews on meta-analysis models, see [2, 5, 6]. Meta-analysis can be applied to various effect sizes collected from individual studies. These include odds ratios and relative risks; standardized mean difference, Cohen's *d*, Hedges' *g*, and Glass's $\Delta$; correlation coefficient and relative metrics;

sensitivity and specificity from diagnostic accuracy studies; and *P*-values. For more comprehensive reviews see Rosenthal [7], Hedges and Olkin [8], and Cooper et al. [9].

## 2. Publication Bias

Publication bias is a threat to any research that attempts to use the published literature, and its potential presence is perhaps the greatest threat to the validity of a meta-analysis [10]. Publication bias exists because research with statistically significant results is more likely to be submitted and published than work with null or nonsignificant results. This issue was memorably termed as the *file-drawer problem* by Rosenthal [11]; nonsignificant results are stored in file drawers without ever being published. In addition to publication bias, other related types of bias exist including pipeline bias, subjective reporting bias, duplicate reporting bias, or language bias (see [12–15] for definitions and examples).

The implication of these various types of bias is that combining only the identified published studies uncritically may lead to an incorrect, usually over optimistic, conclusion [10, 16]. The ability to detect publication bias in a given field is a key strength of meta-analysis because identification of publication bias will challenge the validity of common views in that area and will spur further investigations [17]. There are two types of statistical procedures for dealing with publication bias in meta-analysis: methods for identifying the existence of publication bias and methods for assessing the impact of publications bias [16]. The first includes the funnel plot (and other visualisation methods such as the normal quantile plot) and regression/correlation-based tests, while the second includes the fail-safe (also called file-drawer) number, the trim and fill method, and selection model approaches [10, 14, 18]. Recent approaches include the test for excess significance [19] and the $p$-curve [20].

The most commonly used method is the visual inspection of a funnel plot. This assumes that the results from smaller studies will be more widely spread around the mean effect because of larger random error. The next most frequent method used to assess publication bias is Rosenthal's fail-safe number [21]. Two recent reviews examining the assessment of publication bias in psychology and ecology reported that funnel plots were the most frequently used (24% and 40% resp.), followed by Rosenthal's fail-safe number (22% and 30%, resp.).

### 2.1. Assessing Publication Bias by Computing the Number of Unpublished Studies.
Assessing publication bias can be performed by trying to estimate the number of unpublished studies in the given area a meta-analysis is studying. The fail-safe number represents the number of studies required to refute significant meta-analytic means. Although apparently intuitive, it is in reality difficult to interpret not only because the number of data points (i.e., sample size) for each of $k$ studies is not defined, but also because no benchmarks regarding the fail-safe number exist, unlike Cohen's benchmarks for effect size statistics [22]. However, these versions have been heavily criticised, mainly because such numbers are often misused and misinterpreted [23]. The main reason for the criticism is that, depending on which method is used to estimate the fail-safe $N$, the number of studies can greatly vary.

Although Rosenthal's fail-safe number of publication bias was proposed as early as 1979 and is frequently cited in the literature [11] (over 2000 citations), little attention has been given to the statistical properties of this estimator. This is the aim of the present paper, which is discussed in further detail in Section 3.

Rosenthal [11] introduced what he called the *file drawer problem*. His concern was that some statistically nonsignificant studies may be missing from an analysis (i.e., placed in a file drawer) and that these studies, if included, would nullify the observed effect. By *nullify*, he meant to reduce the effect to a level not statistically significantly different from zero. Rosenthal suggested that rather than speculate on whether the file drawer problem existed, the actual number of studies that would be required to nullify the effect could be calculated [26]. This method calculates the significance of multiple studies by calculating the significance of the mean of the standard normal deviates of each study. Rosenthal's method calculates the number of additional studies $N_R$, with the mean null result necessary to reduce the combined significance to a desired $\alpha$ level (usually 0.05).

The necessary prerequisite is that each study examines a directional null hypothesis such that the effect sizes $\theta_i$ from each study are examined under $\theta_i \leq 0$ or ($\theta_i \geq 0$). Then the null hypothesis of Stouffer [27] test is

$$H_0 : \theta_1 = \cdots = \theta_k = 0. \tag{1}$$

The test statistic for this is

$$Z_S = \frac{\sum_{i=1}^{k} Z_i}{\sqrt{k}}, \tag{2}$$

with $z_i = \theta_i / s_i$, where $s_i$ are the standard errors of $\theta_i$. Under the null hypothesis we have $Z_S \sim N(0, 1)$ [7].

So we get that the number of additional studies $N_R$, with mean null result necessary to reduce the combined significance to a desired $\alpha$ level (usually 0.05 [7, 11]), is found after solving

$$Z_\alpha = \frac{\sum_{i=1}^{k} Z_i}{\sqrt{N_R + k}}. \tag{3}$$

So, $N_R$ is calculated as

$$N_R = \frac{\left( \sum_{i=1}^{k} Z_i \right)^2}{Z_\alpha^2} - k, \tag{4}$$

where $k$ is the number of studies and $Z_\alpha$ is the one-tailed $Z$ score associated with the desired $\alpha$ level of significance. Rosenthal further suggested that if $N_R > 5k + 10$, the likelihood of publication bias would be minimal.

Cooper [28] and Cooper and Rosenthal [29] called this number the fail-safe sample size or fail-safe number. If this number is relatively small, then there is cause for concern. If this number is large, one might be more confident that the effect, although possibly inflated by the exclusion of some studies, is, nevertheless, not zero [30]. This approach is limited in two important ways [26, 31]. First, it assumes that the association in the hidden studies is zero, rather than considering the possibility that some of the studies could have an effect in the reverse direction or an effect that is small but not zero. Therefore, the number of studies required to nullify the effect may be different than the fail-safe number, either larger or smaller. Second, this approach focuses on statistical significance rather than practical or substantive significance (effect sizes). That is, it may allow one to assert that the mean correlation is not zero, but it does not provide an estimate of what the correlation might be (how it has changed in size) after the missing studies are included [23, 32–34]. However, for many fields it remains the gold standard to assess publication bias, since its presentation is conceptually simple and eloquent. In addition, it is computationally easy to perform.

Iyengar and Greenhouse [12] proposed an alternative formula for Rosenthal's fail-safe number, in which the sum of the unpublished studies' standard variates is not zero. In this case the number of unpublished studies $n_\alpha$ is approached through the following equation:

$$Z_\alpha = \frac{\sum_{i=1}^{k} Z_i + n_\alpha M(\alpha)}{\sqrt{n_\alpha + k}}, \quad (5)$$

where $M(\alpha) = -\phi(z_\alpha)/\Phi(z_\alpha)$ (this results immediately from the definition of truncated normal distribution) and $\alpha$ is the desired level of significance. This is justified by the author that the unpublished studies follow a truncated normal distribution with $x \leq z_\alpha$. $\Phi(\cdot)$ and $\phi(\cdot)$ denote the cumulative distribution function (CDF) and probability distribution function (PDF), respectively, of a standard normal distribution.

There are certain other fail-safe numbers which have been described, but their explanation goes beyond the scope of the present article [35]. Duval and Tweedie [36, 37] present three different estimators for the number of missing studies and the method to calculate this has been named Trim and Fill Method. Orwin's [38] approach is very similar to Rosenthal's [11] approach without considering the normal variates but taking Cohen's $d$ [22] to compute a fail-safe number. Rosenberg's fail-safe number is very similar to Rosenthal's and Orwin's fail-safe number [39]. Its difference is that it takes into account the meta-analytic estimate under investigation by incorporating individual weights per study. Gleser and Olkin [40] proposed a model under which the number of unpublished studies in a field where a meta-analysis is undertaken could be estimated. The maximum likelihood estimator of their fail-safe number only needs the number of studies and the maximum $P$ value of the studies. Finally, the Eberly and Casella fail-safe number assumes a Bayesian methodology which aims to estimate the number of unpublished studies in a certain field where a meta-analysis is undertaken [41].

The aim of the present paper is to study the statistical properties of Rosenthal's [11] fail-safe number. In the next section we introduce the statistical theory for computing confidence intervals for Rosenthal's [11] fail-safe number. We initially compute the probability distribution function of $\widehat{N}_R$, which gives formulas for variance and expectation; next, we suggest distributional assumptions for the standard normal variates used in Rosenthal's fail-safe number and finally suggest confidence intervals.

## 3. Statistical Theory

The estimator $\widehat{N}_R$ of unpublished studies is approached through Rosenthal's formula:

$$\widehat{N}_R = \frac{\left(\sum_{i=1}^{k} Z_i\right)^2}{Z_\alpha^2} - k. \quad (6)$$

Let $Z_i$, $i = 1, 2, \ldots, i, \ldots, k$, be i.i.d. random variables with $E[Z_i] = \mu$ and $\mathrm{Var}[Z_i] = \sigma^2$. We discern two cases:

(a) $k$ is fixed or
(b) $k$ is random, assuming additionally that $k \sim \mathrm{Pois}(\lambda)$. This is reasonable since the number of studies

included in a meta-analysis is like observing counts. Other distributions might be assumed, such as the Gamma distribution, but this would require more information or assumptions to compute the parameters of the distribution.

In both cases, estimators of $\mu$, $\sigma^2$, and $\lambda$ can be calculated without distributional assumptions for the $Z_i$ with the method of moments or with distributional assumptions regarding the $Z_i$.

### 3.1. Probability Distribution Function of $\widehat{N}_R$

#### 3.1.1. Fixed k. We compute the PDF of $\widehat{N}_R$ by following the next steps.

Step 1. $Z_1, Z_2, \ldots, Z_i, \ldots, Z_k$ in the formula of the estimator $\widehat{N}_R$ (6) are i.i.d. distributed with $E[Z_i] = \mu$ and $\mathrm{Var}[Z_i] = \sigma^2$. Let $S = \sum_{i=1}^{k} Z_i$ and according to the Lindeberg-Lévy Central Limit Theorem [42], we have

$$\sqrt{k}\left(\frac{S}{k} - \mu\right) \xrightarrow{d} N\left(0, \sigma^2\right) \Longrightarrow S \xrightarrow{d} N\left(k\mu, k\sigma^2\right). \quad (7)$$

So the PDF of $S$ is

$$f_S(s) = \frac{1}{\sqrt{2\pi k}\sigma} \exp\left[-\frac{(s - k\mu)^2}{2k\sigma^2}\right]. \quad (8)$$

Step 2. The PDF of Rosenthal's $\widehat{N}_R$ can be retrieved from a truncated version of (8). From (3), we get that

$$S = Z_\alpha \sqrt{\widehat{N}_R + k}. \quad (9)$$

We advocate that Rosenthal's equations (3) and (9) implicitly impose two conditions which must be taken into account when we seek to estimate the distribution of $N_R$:

$$S \geq 0, \quad (10)$$

$$\widehat{N}_R \geq 0. \quad (11)$$

Expression (10) is justified by the fact that the right hand side of (9) is positive, so then $S \geq 0$. Expression (11) is justified by the fact that $N_R$ expresses the number of studies, so it must be at least 0. Hence, expressions (10) and (11) are satisfied when $S$ is a truncated normal random variable.

The *truncated normal distribution* is a probability distribution related to the normal distribution. Given a normally distributed random variable $X$ with mean $\mu_t$ and variance $\sigma_t^2$, let it be that $X \in (a, b)$, $-\infty \leq a \leq b \leq \infty$. Then $X$ conditional on $a < X < b$ has a truncated normal distribution with PDF: $f_X(x) = (1/\sigma)\phi((x-\mu_t)/\sigma_t)/(\Phi((b-\mu_t)/\sigma_t) - \Phi((a - \mu_t)/\sigma_t))$, for $a \leq x \leq b$ and $f_X(x) = 0$ otherwise [43].

Let it be $S^*$, such that $S^* \geq Z_\alpha \sqrt{k}$. So the PDF of $S^*$ then becomes

$$f_{S^*}(s^*) = \frac{1}{\Phi(\lambda^*)\sqrt{2\pi k}\sigma} \exp\left[-\frac{(s^* - k\mu)^2}{2k\sigma^2}\right], \quad (12)$$

$$s^* \geq Z_\alpha \sqrt{k},$$

where $\lambda^* = (\sqrt{k}\mu - Z_\alpha)/\sigma$.

Then, we have

$$f_{\widehat{N}_R}(n_R) = f_{S^*}(s^*) \left| \frac{dS^*}{dN_R} \right|$$

$$\xrightarrow{(9),(12)} f_{\widetilde{N}_R}(n_R)$$

$$= \frac{Z_\alpha}{2\Phi(\lambda^*)\sqrt{2\pi k\sigma^2(n_R+k)}} \quad (13)$$

$$\times \exp\left[-\frac{\left(Z_\alpha\sqrt{n_R+k}-k\mu\right)^2}{2k\sigma^2}\right], \quad n_R \geq 0.$$

The characteristic function is

$$\psi_{\widehat{N}_R}(t) = E\left[\exp\left(itN_R\right)\right]$$

$$= \frac{\Phi\left((\mu_1+\lambda^*)/\sigma_1\right)}{\Phi(\lambda^*)} \quad (14)$$

$$\cdot \frac{Z_\alpha \exp\left(k^2\mu^2it/\left(Z_\alpha^2-2k\sigma^2it\right)-kit\right)}{\left(Z_\alpha^2-2k\sigma^2it\right)^{1/2}},$$

where $i = \sqrt{-1}$, $\mu_1 = 2k\sqrt{k}\mu\sigma it/(Z_\alpha^2-2k\sigma^2it)$, $\sigma_1 = Z_\alpha^2/(Z_\alpha^2-2k\sigma^2it)$.

From (14) we get

$$E\left[\widehat{N}_R\right] = \frac{k^2\mu^2+k\sigma^2}{Z_\alpha^2} - k + \varepsilon, \quad (15)$$

where $\varepsilon = (\phi(\lambda^*)/\Phi(\lambda^*)) \cdot (k\sigma(\sqrt{k}\mu+Z_\alpha)/Z_\alpha^2)$.

Also,

$$\text{Var}\left[\widehat{N}_R\right] = \frac{2k^2\sigma^2\left(2k\mu^2+\sigma^2\right)}{Z_\alpha^4} + \delta^*, \quad (16)$$

where

$$\delta^* = \frac{\phi(\lambda^*)}{\Phi(\lambda^*)}\left[\frac{k^2\sigma^3\left(5\sqrt{k}\mu+Z_a\right)^2}{Z_\alpha^4}\right.$$

$$\left. - \left(\frac{\phi(\lambda^*)}{\Phi(\lambda^*)}+\lambda^*\right)\frac{k^{3/2}\sigma^2\left(\sqrt{k}\mu+Z_a\right)^2}{Z_\alpha^4}\right]. \quad (17)$$

Proofs for expressions (14), (15), and (16) are given in the Appendix.

*Comments.* Consider the following:

(1) For a significantly large $k$ we have that $\Phi(\lambda^*) \approx 1$. So (13) becomes

$$f_{\widehat{N}_R}(n_R) = \frac{Z_\alpha}{2\sqrt{2\pi k\sigma^2(n_R+k)}}$$

$$\times \exp\left[-\frac{\left(Z_\alpha\sqrt{n_R+k}-k\mu\right)^2}{2k\sigma^2}\right], \quad (18)$$

$$n_R \geq 0.$$

Also we get

$$E\left[\widehat{N}_R\right] = \frac{k^2\mu^2+k\sigma^2}{Z_\alpha^2} - k, \quad (19)$$

$$\text{Var}\left[\widehat{N}_R\right] = \frac{2k^2\sigma^2\left(2k\mu^2+\sigma^2\right)}{Z_\alpha^4}. \quad (20)$$

(2) A limiting element of this computation is that $\widehat{N}_R$ takes discrete values because it describes number of studies, but it has been described by a continuous distribution.

*3.1.2. Random $k$.* It is assumed that $k \sim \text{Pois}(\lambda)$. So taking into account the result from the distribution of $\widehat{N}_R$ for a fixed $k$ we get that the joint distribution of $k$ and $\widehat{N}_R$ is

$$f_{\widehat{N}_R,n}(n_R,k) = f_{\widehat{N}_R}(n_R \mid k=k) \cdot p(k=k)$$

$$\Longrightarrow f_{\widehat{N}_R,n}(n_R,k)$$

$$= \frac{Z_\alpha}{2\Phi(\lambda^*)\sqrt{2\pi k\sigma^2(n_R+k)}}$$

$$\times \exp\left[-\frac{\left(Z_\alpha\sqrt{n_R+k}-k\mu\right)^2}{2k\sigma^2}-\lambda\right] \cdot \frac{\lambda^k}{k!},$$

$$n_R \geq 0, \quad k = 0,1,2,\ldots. \quad (21)$$

*3.2. Expectation and Variance for Rosenthal's Estimator $\widehat{N}_R$*

(a) When $k$ is fixed, expressions (19) and (20) denote the expectation and variance, respectively, for $\widehat{N}_R$. This is derived from the PDF of $\widehat{N}_R$; an additional proof without reference to the PDF is given in the Appendix.

(b) When $k$ is random with $k \sim \text{Pois}(\lambda)$, the expectation and variance of $\widehat{N}_R$ are

$$E\left[\widehat{N}_R\right] = \frac{\lambda^2\mu^2+\lambda\left(\mu^2+\sigma^2\right)}{Z_\alpha^2} - \lambda,$$

$$\text{Var}\left[\widehat{N}_R\right] = \left(\left(4\lambda^3+6\lambda^2+\lambda\right)\mu^4\right.$$

$$+ \left(4\lambda^3+16\lambda^2+6\lambda\right)\mu^2\sigma^2 \quad (22)$$

$$+ \left(2\lambda^2+3\lambda\right)\sigma^4\right)$$

$$\times \left(Z_\alpha^4\right)^{-1} - 2 \cdot \frac{\left(2\lambda^2+\lambda\right)\mu^2+\lambda\sigma^2}{Z_\alpha^2} + \lambda.$$

Proofs are given in the Appendix.

*3.3. Estimators for $\mu$, $\sigma^2$, and $\lambda$.* Having now computed a formula for the variance which is necessary for a confidence interval, we need to estimate $\mu$, $\sigma^2$, and $\lambda$. In both cases, estimators of $\mu$, $\sigma^2$, and $\lambda$ can be calculated without distributional assumptions for the $Z_i$ with the method of moments or with distributional assumptions regarding the $Z_i$.

### 3.3.1. Method of Moments [44]. When $k$ is fixed, we have

$$\widehat{\mu} = \frac{\sum_{i=1}^k \widehat{Z}_i}{k}, \qquad \widehat{\sigma}^2 = \frac{\sum_{i=1}^k \widehat{Z}_i^2}{k} - \left( \frac{\sum_{i=1}^k \widehat{Z}_i}{k} \right)^2. \tag{23}$$

When $k$ is random, we have

$$\widehat{\lambda} = k, \qquad \widehat{\mu} = \frac{\sum_{i=1}^k \widehat{Z}_i}{k},$$

$$\widehat{\sigma}^2 = \frac{\sum_{i=1}^k \widehat{Z}_i^2}{k} - \left( \frac{\sum_{i=1}^k \widehat{Z}_i}{k} \right)^2. \tag{24}$$

### 3.3.2. Distributional Assumptions for the $Z_i$. If we suppose that the $Z_i$ follow a distribution, we would replace the values of $\mu$ and $\sigma^2$ with their distributional values. Below we consider special cases.

*Standard Normal Distribution.* The $Z_i$ follow a standard normal distribution; that is, $Z_i \sim N(0, 1)$. This is the original assumption for the $Z_i$ [11]. In this case we have

$$\widehat{\lambda} = k, \quad \mu = 0, \quad \sigma^2 = 1. \tag{25}$$

Although the origin of the $Z_i$ is from the standard normal distribution, the studies in a meta-analysis are a selected sample of published studies. For this reason, the next distribution is suggested as better.

*Half Normal Distribution.* Here we propose that the $Z_i$ follow a half normal distribution HN(0, 1), which is a special case of folded normal distribution. Before we explain the rational of this distribution, a definition of this type of distribution is provided. A half normal distribution is also a special case of a truncated normal distribution.

*Definition 1.* The *folded normal distribution* is a probability distribution related to the normal distribution. Given a normally distributed random variable $X$ with mean $\mu_f$ and variance $\sigma_f^2$, the random variable $Y = |X|$ has a folded normal distribution [43, 45, 46].

*Remark 2.* The folded normal distribution has the following properties:

(a) probability density function (PDF):

$$f_Y(y) = \frac{1}{\sigma_f \sqrt{2\pi}} \exp \left[ -\frac{\left( -y - \mu_f \right)^2}{2\sigma_f^2} \right]$$

$$+ \frac{1}{\sigma_f \sqrt{2\pi}} \exp \left[ -\frac{\left( y - \mu_f \right)^2}{2\sigma_f^2} \right], \tag{26}$$

$$\text{for } y \geq 0,$$

(b)

$$E[Y] = \sigma_f \sqrt{\frac{2}{\pi}} \exp \left( \frac{-\mu_f^2}{2\sigma_f^2} \right) + \mu_f \left[ 1 - 2\Phi \left( \frac{-\mu_f}{\sigma_f} \right) \right],$$

$$\text{Var}[Y]$$

$$= \mu_f^2 + \sigma_f^2$$

$$- \left\{ \sigma_f \sqrt{\frac{2}{\pi}} \exp \left( \frac{-\mu_f^2}{2\sigma_f^2} \right) + \mu_f \left[ 1 - 2\Phi \left( \frac{-\mu_f}{\sigma_f} \right) \right] \right\}^2. \tag{27}$$

*Remark 3.* When $\mu_f = 0$, the distribution of $Y$ is a *half normal distribution*. This distribution is identical to the *truncated normal distribution*, with left truncation point 0 and no right truncation point. For this distribution we have the following.

(a) $f_Y(y) = (\sqrt{2}/\sigma_f \sqrt{\pi}) \exp(-y^2/2\sigma_f^2)$, for $y \geq 0$.

(b) $E[Y] = \sigma_f \sqrt{2/\pi}$, $\text{Var}[Y] = \sigma_f^2(1 - 2/\pi)$.

*Assumption 4.* The $Z_i$ in Rosenthal's estimator $N_R$ are derived from a half normal distribution, based on a normal distribution $N(0, 1)$.

*Support.* When a researcher begins to perform a meta-analysis, the sample of studies is drawn from those studies that are already published. So his sample is most likely biased by some sort of selection bias, produced via a specific selection process [47]. Thus, although when we study Rosenthal's $N_R$ assuming that all $Z_i$ are drawn from the normal distribution, they are in essence drawn from a truncated normal distribution. This has been commented on by Iyengar and Greenhouse [12] and Schonemann and Scargle [48]. But at which point is this distribution truncated? We would like to advocate that the half normal distribution, based on a normal distribution $N(0, 1)$, is the one best representing the $Z_i$ Rosenthal uses to compute his fail-safe $N_R$. The reasons for this are as follows.

(1) Firstly, assuming that all $Z_i$ are of the same sign does not impede the significance of the results from each study. That is, the test is significant when either $Z_i > Z_{\alpha/2}$ or $Z_i < Z_{1-\alpha/2}$ occurs.

(2) However, when a researcher begins to perform a meta-analysis of studies, many times $Z_i$ can be either positive or negative. Although this is true, when the researcher is interested in doing a meta-analysis, usually the $Z_i$ that have been published are indicative of a significant effect of the same direction (thus $Z_i$ have the same sign) or are at least indicative of such an association without being statistically significant, thus producing $Z_i$ of the same sign but not producing significance (e.g., the confidence interval of the effect might include the null value).

(3) There will definitely be studies that produce a totally opposite effect, thus producing an effect of opposite

direction, but these will definitely be a minority of the studies. Also there is the case that these other signed $Z_i$ are not significant.

Hence, in this case

$$\widehat{\lambda} = k, \qquad \mu = \sqrt{\frac{2}{\pi}}, \qquad \sigma^2 = 1 - \frac{2}{\pi}. \tag{28}$$

*Skew Normal Distribution.* Here we propose that the $Z_i$ follow a skew normal distribution; that is, $Z_i \sim \text{SN}(\xi, \omega, \alpha)$.

*Definition 5.* The skew normal distribution is a continuous probability distribution that generalises the normal distribution to allow for nonzero skewness. A random variable $X$ follows a univariate skew normal distribution with location parameter $\xi \in R$, scale parameter $\omega \in R^+$, and skewness parameter $\alpha \in R$ [49], if it has the density

$$f_X(x) = \frac{2}{\omega} \phi\left(\frac{x - \xi}{\omega}\right) \Phi\left(\alpha \frac{x - \xi}{\omega}\right) \quad x \in R. \tag{29}$$

Note that if $\alpha = 0$, the density of $X$ reduces to the $N(\xi, \omega^2)$.

*Remark 1.* The expectation and variance of $X$ are [49]

$$E[X] = \xi + \omega\delta\sqrt{\frac{2}{\pi}}, \quad \text{where } \delta = \frac{\alpha}{\sqrt{1 + \alpha^2}},$$
$$\text{Var}(X) = \omega^2 \left(1 - \frac{2\delta^2}{\pi}\right). \tag{30}$$

*Remark 2.* The methods of moments estimators for $\xi, \omega$, and $\delta$ are [50, 51]

$$\widetilde{\xi} = m_1 - a_1 \left(\frac{m_3}{b_1}\right)^{1/3}, \qquad \widetilde{\omega}^2 = m_2 - a_1^2 \left(\frac{m_3}{b_1}\right)^{2/3},$$
$$\widetilde{\delta} = \left[a_1^2 + m_2 \left(\frac{b_1}{m_3}\right)^{2/3}\right]^{-1/2}, \tag{31}$$

where $a_1 = \sqrt{2/\pi}$, $b_1 = (4/\pi - 1)a_1$, $m_1 = n^{-1}\sum_{i=1}^n X_i$, $m_2 = n^{-1}\sum_{i=1}^n (X_i - m_1)^2$, and $m_3 = n^{-1}\sum_{i=1}^n (X_i - m_1)^3$. The sign of $\widetilde{\delta}$ is taken to be the sign of $m_3$.

*Explanation.* The skew normal distribution allows for a dynamic way to fit the available $Z$-scores. The fact that there is ambiguity on the derivation of the standard deviates from each study from a normal or a truncated normal distribution creates the possibility that the distribution could be a skew normal, with the skewness being attributed that we are including only the published $Z$-scores in the estimation of Rosenthal's [11] estimator.

Hence, in this case and taking the method of moments estimators of $\xi, \omega$, and $\delta$, we get

$$\widehat{\lambda} = n, \qquad \widehat{\mu} = \widetilde{\xi} + \widetilde{\omega}\widetilde{\delta}\sqrt{\frac{2}{\pi}}, \qquad \widehat{\sigma}^2 = \widetilde{\omega}^2 \left(1 - \frac{2\widetilde{\delta}^2}{\pi}\right), \tag{32}$$

where $a_1 = \sqrt{2/\pi}$, $b_1 = (4/\pi - 1)$, $m_1 = n^{-1}\sum_{i=1}^n Z_i$, $m_2 = n^{-1}\sum_{i=1}^n (Z_i - m_1)^2$, and $m_3 = n^{-1}\sum_{i=1}^n (Z_i - m_1)^3$.

### 3.4. Methods for Confidence Intervals

*3.4.1. Normal Approximation.* In the previous section, formulas for computing the variance of $\widehat{N}_R$ were derived. We compute asymptotic $(1 - \alpha/2)\%$ confidence intervals for $N_R$ as

$$\left(\widehat{N}_{R_{\text{low}}}, \widehat{N}_{R_{\text{up}}}\right)$$
$$= \left(\widehat{N}_R - Z_{1-\alpha/2}\sqrt{\widehat{\text{Var}}\left[\widehat{N}_R\right]}, \widehat{N}_R + Z_{1-\alpha/2}\sqrt{\widehat{\text{Var}}\left[\widehat{N}_R\right]}\right), \tag{33}$$

where $Z_{1-\alpha/2}$ is the $(1-\alpha/2)$th quantile of the standard normal distribution.

The variance of $\widehat{N}_R$ for a given set of values $Z_i$ depends firstly on whether the number of studies $k$ is fixed or random and secondly on whether the estimators of $\mu$, $\sigma^2$, and $\lambda$ are derived from the method of moments or from the distributional assumptions.

*3.4.2. Nonparametric Bootstrap.* Bootstrap is a well-known resampling methodology for obtaining nonparametric confidence intervals of a parameter [52, 53]. In most statistical problems one needs an estimator of a parameter of interest as well as some assessment of its variability. In many such problems the estimators are complicated functionals of the empirical distribution function and it is difficult to derive trustworthy analytical variance estimates for them. The primary objective of this technique is to estimate the sampling distribution of a statistic. Essentially, bootstrap is a method that mimics the process of sampling from a population, like one does in Monte Carlo simulations, but instead drawing samples from the observed sampling data. The tool of this mimic process is the Monte Carlo algorithm of Efron [54]. This process is explained properly by Efron and Tibshirani [55] and Davison and Hinkley [56], who also noted that bootstrap confidence intervals are approximate, yet better than the standard ones. Nevertheless, they do not try to replace the theoretical ones and bootstrap is not a substitute for precise parametric results, but rather a way to reasonably proceed when such results are unavailable.

Nonparametric resampling makes no assumptions concerning the distribution of, or model for, the data [57]. Our data is assumed to be a vector $\mathbf{Z}_{\text{obs}}$ of $k$ independent observations, and we are interested in a confidence interval for $\widehat{\theta}(\mathbf{Z}_{\text{obs}})$. The general algorithm for a nonparametric bootstrap is as follows.

(1) Sample $k$ observations randomly with replacement from $\mathbf{Z}_{\text{obs}}$ to obtain a bootstrap data set, denoted by $\mathbf{Z}^*$.

(2) Calculate the bootstrap version of the statistic of interest $\widehat{\theta}^* = \widehat{\theta}(\mathbf{Z}^*)$.

(3) Repeat steps (1) and (2) several times, say $B$, to obtain an estimate of the bootstrap distribution.

In our case

(1) compute a random sample from the initial sample of $Z_i$, size $k$,

(2) compute $N_R^*$ from this sample,

(3) repeat these processes $b$ times.

Then the bootstrap estimator of $N_R$ is

$$N_{R.\text{bootstrap}} = \frac{\sum N_R^*}{b}. \tag{34}$$

From this we can compute also confidence intervals for $N_{R.\text{bootstrap}}$.

In the next section, we investigate these theoretical aspects with simulations and examples.

## 4. Simulations and Results

The method for simulations is as follows.

(1) Initially we draw random numbers from the following distributions:

    (a) standard normal distribution,

    (b) half normal distribution $(0, 1)$,

    (c) skew normal distribution with negative skewness $SN(\delta = -0.5, \xi = 0, \omega = 1)$,

    (d) skew normal distribution with positive skewness $SN(\delta = 0.5, \xi = 0, \omega = 1)$.

(2) The numbers drawn from each distribution represent the number of studies in a meta-analysis and we have chosen $k = 5, 15, 30$, and $50$. When $k$ is assumed to be random, then the parameter $\lambda$ is equal to the values chosen for the simulation, that is, $5, 15, 30$, and $50$, respectively.

(3) We compute the normal approximation confidence interval with the formulas described in Section 3 and the bootstrap confidence interval. We also discern whether the number of studies is fixed or random. For the computation of the bootstrap confidence interval, we generate 1,000 bootstrap samples each time. We also study the performance of the different distributional estimators in cases where the distributional assumption is not met, thus comparing each of the six confidence interval estimators under all four distributions.

(4) We compute the coverage probability comparing with the true value of Rosenthal's fail-safe number. When the number of studies is fixed the true value of Rosenthal's number is

$$E\left[\widehat{N}_R\right] = \frac{k^2\mu^2 + k\sigma^2}{Z_\alpha^2} - k. \tag{35}$$

When the number of studies is random [from a Poisson $(\lambda)$ distribution] the true value of Rosenthal's number is

$$E\left[\widehat{N}_R\right] = \frac{\lambda^2\mu^2 + \lambda\left(\mu^2 + \sigma^2\right)}{Z_\alpha^2} - \lambda. \tag{36}$$

We execute the above procedure 10,000 times each time. Our alpha-level is considered 5%.

This process is shown schematically in Table 1. All simulations were performed in $R$ and the code is shown in the Supplementary Materials (see Supplementary Materials available online at http://dx.doi.org/10.1155/2014/825383).

We observe from Table 2 and Figure 1 that the bootstrap confidence intervals perform the poorest both when the number of studies is considered fixed or random. The only case in which they perform acceptably is when the distribution is half normal and the number of studies is fixed. The moment estimators of variance perform either poorly or too efficiently in all cases, with coverages being under 90% or near 100%. The most acceptable confidence intervals for Rosenthal's estimator appear to be in the distribution-based method and to be much better for a fixed number of studies than for random number of studies. We also observe that, for the distribution-based confidence intervals in the fixed category, the half normal distribution HN(0,1) produces coverages which are all 95%. This is also stable for all number of studies in a meta-analysis. When the distributional assumption is not met the coverage is poor except for the cases of the positive and negative skewness skew normal distributions which perform similarly, possibly due to symmetry.

In the next sections, we give certain examples and we present the lower limits of confidence intervals for testing whether $N_R > 5k + 10$, according to the suggested rule of thumb by [11]. We choose only the variance from a fixed number of studies when the $Z_i$ are drawn from a half normal distribution HN(0,1).

## 5. Examples

In this section, we present two examples of meta-analyses from the literature. The first study is a meta-analysis of the effect of probiotics for preventing antibiotic-associated diarrhoea and included 63 studies [24]. The second meta-analysis comes from the psychological literature and is a meta-analysis examining reward, cooperation, and punishment, including analysis of 148 effect sizes [25]. For each meta-analysis, we computed Rosenthal's fail-safe number and the respective confidence interval with the methods described above (Table 3).

We observe that both fail-safe numbers exceed Rosenthal's rule of thumb, but some lower confidence intervals, especially in the first example, go as low as 369, which only slightly surpasses the rule of thumb ($5 * 63 + 10 = 325$ in this case). This is not the case with the second example. Hence the confidence interval, especially the lower confidence interval value, is important to establish whether the fail-safe number surpasses the rule of thumb.

In the next section, we present a table with values according to which future researchers can get advice on whether their value truly supersedes the rule of thumb.

## 6. Suggested Confidence Limits for $N_R$

We wish to answer the question whether $N_R > 5k + 10$ for a given level of significance and the estimator $\widehat{N}_R$, which is

TABLE 1: Schematic table for simulation plan.

| | Variance formula for normal approximation confidence intervals | | Bootstrap confidence interval | Real value of $N_R$ |
|---|---|---|---|---|
| | Distributional | Moments | | |
| $k = 5, 15, 30, 50$ studies<br>Draw $Z_i$ from N(0,1),<br>HN(0,1),<br>SN($\delta = -0.5, \xi = 0, \omega = 1$),<br>SN($\delta = 0.5, \xi = 0, \omega = 1$) | Fixed $k$ (standard normal distribution values $\mu = 0, \sigma^2 = 1$; half normal distribution $\mu = \sqrt{2/\pi}, \sigma^2 = 1 - 2/\pi$; skew normal with negative skewness $\mu = -\sqrt{1/2\pi}, \sigma^2 = 1 - 1/(2\pi)$; skew normal with positive skewness $\mu = \sqrt{1/2\pi}, \sigma^2 = 1 - 1/(2\pi)$)<br><br>$\mathrm{Var}\left[\hat{N}_R\right] = \dfrac{2k^2\sigma^2\left(2k\mu^2 + \sigma^2\right)}{Z_\alpha^4}$ | Fixed $k\left(\hat{\mu} = \dfrac{\sum_{i=1}^k \overline{Z_i}}{k}, \hat{\sigma}^2 = \dfrac{\sum_{i=1}^k \overline{Z_i^2}}{k} - \left(\dfrac{\sum_{i=1}^k \overline{Z_i}}{k}\right)^2\right)$<br><br>$\mathrm{Var}\left[\hat{N}_R\right] = \dfrac{2k^2\hat{\sigma}^2\left(2k\hat{\mu}^2 + \hat{\sigma}^2\right)}{Z_\alpha^4}$ | Using the $N_R^*$ we compute the $N_{R,\text{bootstrap}}$ and respectively the standard error needed to compute the confidence interval | Fixed $k$ (standard normal distribution values $\mu = 0, \sigma^2 = 1$; half normal distribution $\mu = \sqrt{2/\pi}, \sigma^2 = 1 - 2/\pi$; skew normal with negative skewness $\mu = -\sqrt{1/2\pi}, \sigma^2 = 1 - 1/(2\pi)$; skew normal with positive skewness $\mu = \sqrt{1/2\pi}, \sigma^2 = 1 - 1/(2\pi)$)<br><br>$E\left[\hat{N}_R\right] = \dfrac{k^2\mu^2 + k\sigma^2}{Z_\alpha^2} - k$ |
| | Random $k$ (standard normal distribution values $\mu = 0, \sigma^2 = 1$; half normal distribution $\mu = \sqrt{2/\pi}, \sigma^2 = 1 - 2/\pi$; skew normal with negative skewness $\mu = -\sqrt{1/2\pi}, \sigma^2 = 1 - 1/(2\pi)$; skew normal with positive skewness $\mu = \sqrt{1/2\pi}, \sigma^2 = 1 - 1/(2\pi); \lambda = 5, 15, 30, 50$)<br><br>$\mathrm{Var}\left[\hat{N}_R\right] = \dfrac{\left(4\lambda^3 + 6\lambda^2 + \lambda\right)\mu^4 + \left(4\lambda^3 + 16\lambda^2 + 6\lambda\right)\mu^2\sigma^2}{Z_\alpha^4}$<br>$+ \dfrac{\left(2\lambda^2 + 3\lambda\right)\sigma^4}{Z_\alpha^4} - 2 \cdot \dfrac{\left(2\lambda^2 + \lambda\right)\mu^2 + \lambda\sigma^2}{Z_\alpha^2} + \lambda$ | Random $k\left(\hat{\mu} = \dfrac{\sum_{i=1}^k \overline{Z_i}}{k}, \hat{\sigma}^2 = \dfrac{\sum_{i=1}^k \overline{Z_i^2}}{k} - \left(\dfrac{\sum_{i=1}^k \overline{Z_i}}{k}\right)^2\right),$<br>$\lambda = 5, 15, 30, 50$<br><br>$\mathrm{Var}\left[\hat{N}_R\right] = \dfrac{\left(4\lambda^3 + 6\lambda^2 + \lambda\right)\hat{\mu}^4 + \left(4\lambda^3 + 16\lambda^2 + 6\lambda\right)\hat{\mu}^2\hat{\sigma}^2}{Z_\alpha^4}$<br>$+ \dfrac{\left(2\lambda^2 + 3\lambda\right)\hat{\sigma}^4}{Z_\alpha^4} - 2 \cdot \dfrac{\left(2\lambda^2 + \lambda\right)\hat{\mu}^2 + \lambda\hat{\sigma}^2}{Z_\alpha^2} + \lambda$ | Using the $N_R^*$ we compute the $N_{R,\text{bootstrap}}$ and respectively the standard error needed to compute the confidence interval | Random $k$ (standard normal distribution values $\mu = 0, \sigma^2 = 1$; half normal distribution $\mu = \sqrt{2/\pi}, \sigma^2 = 1 - 2/\pi$; skew normal with negative skewness $\mu = -\sqrt{1/2\pi}, \sigma^2 = 1 - 1/(2\pi)$; skew normal with positive skewness $\mu = \sqrt{1/2\pi}, \sigma^2 = 1 - 1/(2\pi); \lambda = 5, 15, 30, 50$)<br><br>$E\left[\hat{N}_R\right] = \dfrac{\lambda^2\mu^4 + \lambda\left(\mu^2 + \sigma^2\right)}{Z_\alpha^2} - \lambda$ |

TABLE 2: Probability coverage of the different methods for confidence intervals (CI) according to the number of studies $k$. The figure is organised as follows: the $Z_i$ are drawn from four different distributions (standard normal distribution, half normal distribution, skew normal distribution with negative skewness, and skew normal with positive skewness).

| Draw $Z_i$ from | | | Values of $\mu$ and $\sigma^2$ from the standard normal distribution | | | | Values of $\mu$ and $\sigma^2$ from the half normal distribution HN(0, 1) | | | | Values of $\mu$ and $\sigma^2$ from the skew normal distribution with negative skewness SN($\delta = -0.5, \xi = 0, \omega = 1$) | | | | Values of $\mu$ and $\sigma^2$ from the skew normal distribution with positive skewness SN($\delta = 0.5, \xi = 0, \omega = 1$) | | | |
|---|---|---|---|---|---|---|---|---|---|---|---|---|---|---|---|---|---|---|
| | | | $k = 5$ | $k = 15$ | $k = 30$ | $k = 50$ | $k = 5$ | $k = 15$ | $k = 30$ | $k = 50$ | $k = 5$ | $k = 15$ | $k = 30$ | $k = 50$ | $k = 5$ | $k = 15$ | $k = 30$ | $k = 50$ |
| Standard normal distribution | Fixed $k$ | Distribution Based CI | 0.948 | 0.950 | 0.948 | 0.952 | 0.994 | 0.110 | 0.002 | 0.000 | 0.985 | 0.999 | 1.000 | 1.000 | 0.982 | 0.998 | 1.000 | 1.000 |
| | | Moments Based CI | 0.933 | 0.996 | 1.000 | 1.000 | 0.529 | 0.088 | 0.005 | 0.000 | 0.842 | 0.686 | 0.337 | 0.120 | 0.842 | 0.686 | 0.337 | 0.120 |
| | | Bootstrap CI | 0.929 | 0.996 | 1.000 | 1.000 | 0.534 | 0.089 | 0.005 | 0.000 | 0.830 | 0.680 | 0.337 | 0.120 | 0.830 | 0.680 | 0.337 | 0.120 |
| | Random $k$ | Distribution Based CI | 0.966 | 0.956 | 0.951 | 0.955 | 0.999 | 1.000 | 0.084 | 0.001 | 0.998 | 1.000 | 1.000 | 1.000 | 0.990 | 0.999 | 1.000 | 1.000 |
| | | Moments Based CI | 1.000 | 1.000 | 1.000 | 1.000 | 0.535 | 0.094 | 0.006 | 0.000 | 1.000 | 0.702 | 0.338 | 0.122 | 1.000 | 0.702 | 0.338 | 0.122 |
| | | Bootstrap CI | 0.929 | 0.996 | 1.000 | 1.000 | 0.429 | 0.074 | 0.004 | 0.000 | 0.804 | 0.649 | 0.322 | 0.115 | 0.804 | 0.649 | 0.322 | 0.115 |
| Half normal distribution HN(0, 1) | Fixed $k$ | Distribution Based CI | 0.635 | 0.021 | 0.000 | 0.000 | 0.945 | 0.952 | 0.951 | 0.948 | 0.864 | 0.624 | 0.279 | 0.053 | 0.841 | 0.483 | 0.142 | 0.014 |
| | | Moments Based CI | 0.861 | 0.187 | 0.000 | 0.000 | 0.771 | 0.880 | 0.911 | 0.927 | 0.885 | 0.657 | 0.126 | 0.003 | 0.885 | 0.657 | 0.126 | 0.003 |
| | | Bootstrap CI | 0.858 | 0.217 | 0.000 | 0.000 | 0.775 | 0.884 | 0.913 | 0.929 | 0.887 | 0.672 | 0.138 | 0.003 | 0.887 | 0.672 | 0.138 | 0.003 |
| | Random $k$ | Distribution Based CI | 0.720 | 0.027 | 0.000 | 0.000 | 0.989 | 0.995 | 0.996 | 0.997 | 0.966 | 0.915 | 0.762 | 0.459 | 0.901 | 0.578 | 0.198 | 0.027 |
| | | Moments Based CI | 1.000 | 1.000 | 0.130 | 0.000 | 0.806 | 0.937 | 0.971 | 0.984 | 1.000 | 1.000 | 0.995 | 0.358 | 1.000 | 1.000 | 0.995 | 0.358 |
| | | Bootstrap CI | 0.858 | 0.217 | 0.000 | 0.000 | 0.715 | 0.859 | 0.899 | 0.920 | 0.885 | 0.698 | 0.152 | 0.004 | 0.885 | 0.698 | 0.152 | 0.004 |
| Skew normal distribution with negative skewness SN($\delta = -0.5, \xi = 0, \omega = 1$) | Fixed $k$ | Distribution Based CI | 0.872 | 0.666 | 0.399 | 0.174 | 0.980 | 0.472 | 0.184 | 0.048 | 0.953 | 0.970 | 0.977 | 0.981 | 0.944 | 0.948 | 0.949 | 0.957 |
| | | Moments Based CI | 0.917 | 0.979 | 0.944 | 0.858 | 0.597 | 0.375 | 0.200 | 0.074 | 0.845 | 0.860 | 0.882 | 0.895 | 0.845 | 0.860 | 0.882 | 0.895 |
| | | Bootstrap CI | 0.912 | 0.978 | 0.945 | 0.857 | 0.586 | 0.377 | 0.199 | 0.074 | 0.840 | 0.857 | 0.881 | 0.894 | 0.840 | 0.857 | 0.881 | 0.894 |
| | Random $k$ | Distribution Based CI | 0.903 | 0.688 | 0.409 | 0.178 | 0.996 | 1.000 | 0.687 | 0.306 | 0.987 | 0.997 | 0.998 | 0.999 | 0.965 | 0.964 | 0.965 | 0.968 |
| | | Moments Based CI | 1.000 | 1.000 | 0.999 | 0.968 | 0.609 | 0.399 | 0.237 | 0.103 | 1.000 | 0.872 | 0.886 | 0.902 | 1.000 | 0.872 | 0.886 | 0.902 |
| | | Bootstrap CI | 0.912 | 0.978 | 0.945 | 0.857 | 0.534 | 0.342 | 0.181 | 0.066 | 0.818 | 0.845 | 0.874 | 0.889 | 0.818 | 0.845 | 0.874 | 0.889 |
| Skew normal distribution with positive skewness SN($\delta = 0.5, \xi = 0, \omega = 1$) | Fixed $k$ | Distribution Based CI | 0.880 | 0.673 | 0.402 | 0.164 | 0.982 | 0.471 | 0.186 | 0.050 | 0.956 | 0.972 | 0.976 | 0.979 | 0.948 | 0.952 | 0.951 | 0.955 |
| | | Moments Based CI | 0.923 | 0.980 | 0.947 | 0.852 | 0.596 | 0.372 | 0.201 | 0.076 | 0.850 | 0.865 | 0.874 | 0.896 | 0.850 | 0.865 | 0.874 | 0.896 |
| | | Bootstrap CI | 0.918 | 0.978 | 0.946 | 0.846 | 0.583 | 0.372 | 0.200 | 0.077 | 0.841 | 0.862 | 0.873 | 0.896 | 0.841 | 0.862 | 0.873 | 0.896 |
| | Random $k$ | Distribution Based CI | 0.911 | 0.696 | 0.415 | 0.169 | 0.996 | 1.000 | 0.683 | 0.314 | 0.989 | 0.996 | 0.998 | 0.999 | 0.967 | 0.967 | 0.964 | 0.966 |
| | | Moments Based CI | 1.000 | 1.000 | 0.999 | 0.964 | 0.606 | 0.399 | 0.236 | 0.105 | 1.000 | 0.875 | 0.880 | 0.905 | 1.000 | 0.875 | 0.880 | 0.905 |
| | | Bootstrap CI | 0.918 | 0.978 | 0.946 | 0.846 | 0.534 | 0.335 | 0.180 | 0.068 | 0.819 | 0.850 | 0.868 | 0.893 | 0.819 | 0.850 | 0.868 | 0.893 |

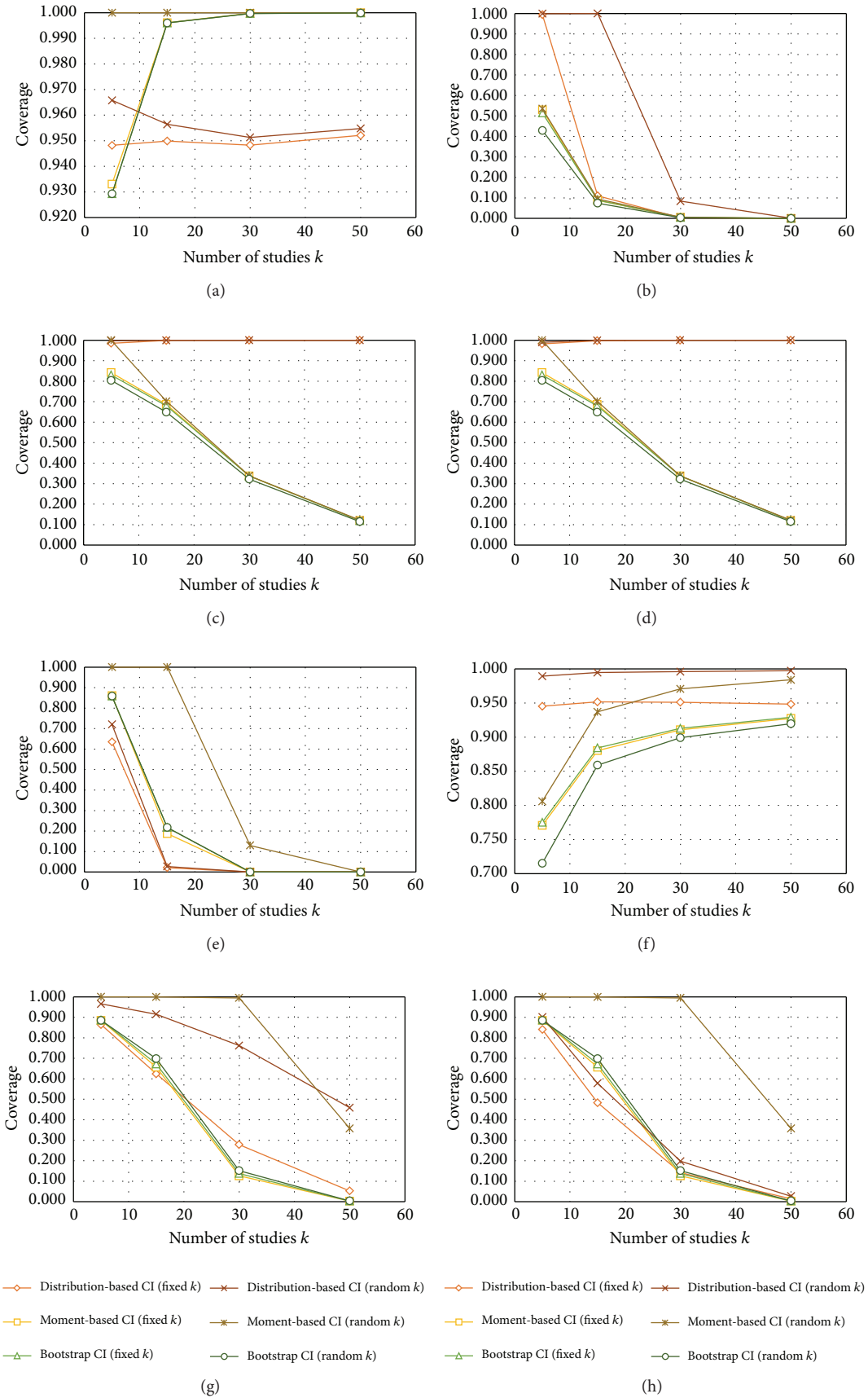

FIGURE 1: Continued.

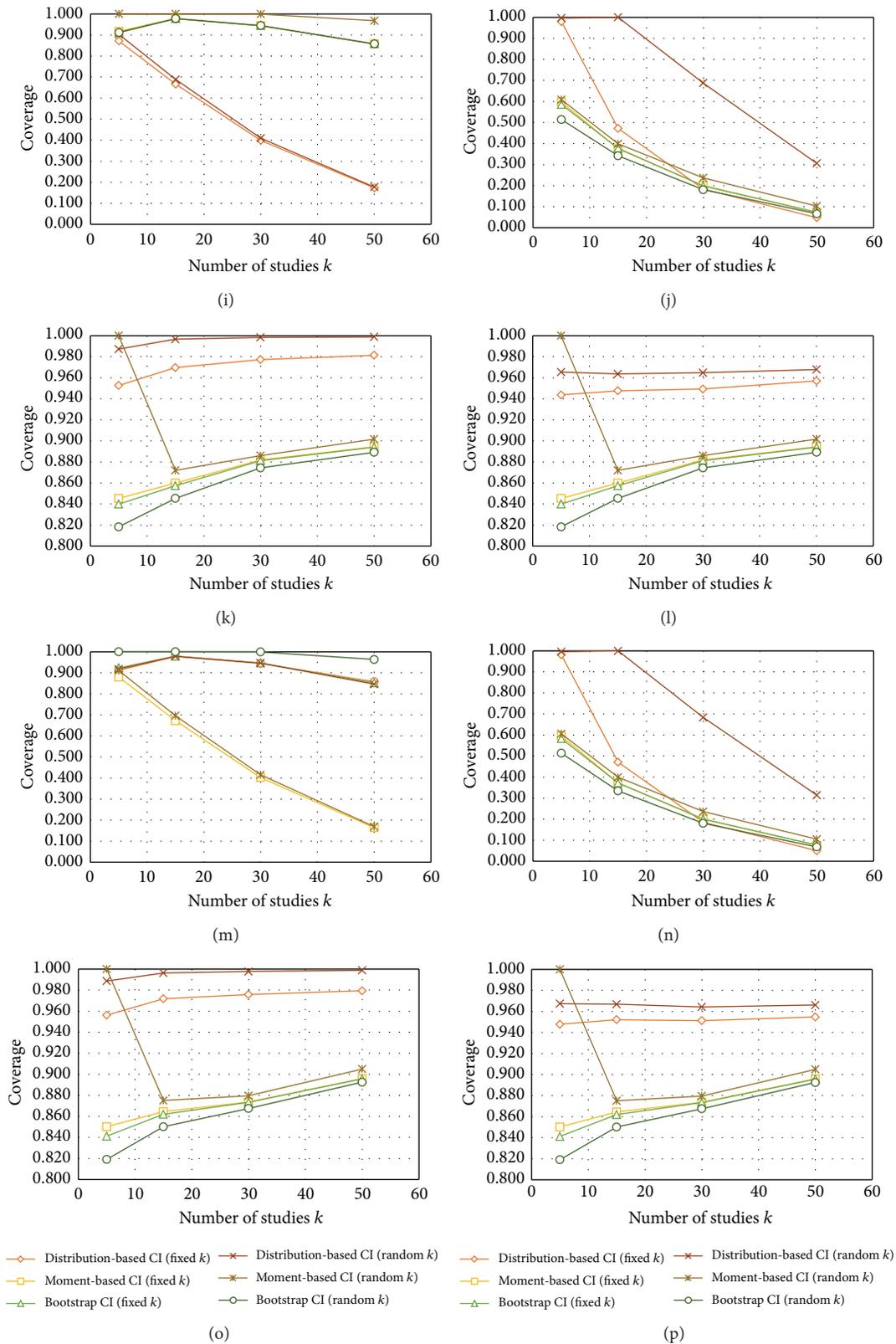

Figure 1: This figures shows the probability coverage of the different methods for confidence intervals (CI) according to the number of studies $k$. The figure is organised as follows: the $Z_i$ are drawn from four different distributions (standard normal distribution, half normal distribution, skew normal with negative skewness, and skew normal with positive skewness) which are depicted in ((a)–(d), (e)–(h), (i)–(l), and (m)–(p)), respectively. The different values of $\mu$ and $\sigma^2$ for the variance correspond to the standard normal distribution ((a), (e), (i), and (m)), half normal distribution ((b), (f), (j), and (n)), skew normal with negative skewness ((c), (g), (k), and (o)), and skew normal with positive skewness ((d), (h), (l), and (p)).

TABLE 3: Confidence intervals for the examples of meta-analyses.

| | Fixed number of studies | | Random number of studies | | |
| | Distribution based CI | Moment based CI | Distribution based CI | Moment based CI | Bootstrap based CI |
|---|---|---|---|---|---|
| Study 1 [24] Rosenthal's $N_R = 2124$ | (2060, 2188) | (788, 3460) | (2059, 2189) | (369, 3879) | (740, 3508) |
| Study 2 [25] Rosenthal's $N_R = 73860$ | (73709, 74012) | (51618, 96102) | (73707, 74013) | (40976, 106745) | (51662, 96059) |

the rule of thumb suggested by Rosenthal. We formulate a hypothesis test according to which

$$H_0 : N_R \leq 5k + 10$$
$$H_1 : N_R > 5k + 10. \tag{37}$$

An asymptotic test statistic for this is

$$T \& = \frac{\widehat{N}_R - 5k - 10}{\sqrt{\text{Var}\left[\widehat{N}_R\right]}} \xrightarrow{d} N(0, 1), \tag{38}$$

under the null hypothesis.

So we reject the null hypothesis if $(\widehat{N}_R - 5k - 10)/\sqrt{\text{Var}[\widehat{N}_R]} > Z_\alpha \Rightarrow \widehat{N}_R > Z_\alpha \sqrt{\text{Var}[\widehat{N}_R]} + 5k + 10$.

In Table 4 we give the limits of $N_R$ above which we are 95% confident that $N_R > 5k + 10$. For example if a researcher performs a meta-analysis of 25 studies, the rule of thumb suggests that over $5 \cdot 25 + 10 = 135$ studies there is no publication bias. The present approach and the values of Table 4 suggest that we are 95% confident for this when $N_R$ exceeds 209 studies. So this approach allows for inferences about Rosenthal's $\widehat{N}_R$ and is also slightly more conservative especially when Rosenthal's fail-safe number is characterised from overestimating the number of published studies.

## 7. Discussion and Conclusion

The purpose of the present paper was to assess the efficacy of confidence intervals for Rosenthal's fail-safe number. We initially defined publication bias and described an overview of the available literature on fail-safe calculations in meta-analysis. Although Rosenthal's estimator is highly used by researchers, its properties and usefulness have been questioned [48, 58].

The original contributions of the present paper are its theoretical and empirical results. First, we developed statistical theory allowing us to produce confidence intervals for Rosenthal's fail-safe number. This was produced by discerning whether the number of studies analysed in a meta-analysis is fixed or random. Each case produces different variance estimators. For a given number of studies and a given distribution, we provided five variance estimators: moment- and distribution-based estimators based on whether the number of studies is fixed or random and on bootstrap confidence intervals. Secondly, we examined four distributions by which we can simulate and test our hypotheses of variance, namely, standard normal distribution, half normal distribution, a positive skew normal distribution, and a negative skew normal distribution. These four distributions were chosen as closest to the nature of the $Z_i$s. The half normal distribution variance estimator appears to present the best coverage for the confidence intervals. Hence, this might support the hypothesis that $Z_i$s are derived from a half normal distribution. Thirdly, we provide a table of lower confidence intervals for Rosenthal's estimator.

The limitations of the study initially stem from the flaws associated with Rosenthal's estimator. This usually means that the number of negative studies needed to disprove the result is highly overestimated. However, its magnitude can give an indication for no publication bias. Another possible flaw could come from the simulation planning. We could try more values for the skew normal distribution, for which we tried only two values in present paper.

The implications of this research for applied researchers in psychology, medicine, and social sciences, which are the fields that predominantly use Rosenthal's fail-safe number, are immediate. Table 4 provides an accessible reference for researchers to consult and apply this more conservative rule for Rosenthal's number. Secondly, the formulas for the variance estimator are all available to researchers, so they can compute normal approximation confidence intervals on their own. The future step that needs to be attempted is to develop an R-package program or a Stata program to execute this quickly and efficiently and make it available to the public domain. This will allow widespread use of these techniques.

In conclusion, the present study is the first in the literature to study the statistical properties of Rosenthal's fail-safe number. Statistical theory and simulations were presented and tables for applied researchers were also provided. Despite the limitations of Rosenthal's fail-safe number, it can be a trustworthy way to assess publication bias, especially under the more conservative nature of the present paper.

## Appendices

## A. Proofs for Expressions (19), (20), and (22)

*(a) Fixed k.* $Z_1, Z_2, \ldots, Z_i, \ldots, Z_k$ in the formula of the estimator $\widehat{N}_R$ (9) are i.i.d. distributed with $E[Z_i] = \mu$ and $\text{Var}[Z_i] = \sigma^2$. Let $S = \sum_{i=1}^{k} Z_i$; then, according to the Lindeberg-Lévy Central Limit Theorem [42], we have

$$\sqrt{k}\left(\frac{S}{k} - \mu\right) \xrightarrow{d} N(0, \sigma^2) \Longrightarrow S \xrightarrow{d} N(k\mu, k\sigma^2). \quad \text{(A.1)}$$

Table 4: 95% one-sided confidence limits above which the estimated $N_R$ is significantly higher than $5k + 10$, which is the rule of thumb suggested by Rosenthal [11]. $k$ represents the number of studies included in a meta-analysis. We choose the variance from a fixed number of studies when the $Z_i$ are drawn from a half normal distribution $HN(0, 1)$, as this performed best in the simulations.

| $k$ | Cutoff point | $k$ | Cutoff point | $k$ | Cutoff point | $k$ | Cutoff point |
|-----|------|-----|------|-----|------|-----|------|
| 1 | 17 | 41 | 369 | 81 | 842 | 121 | 1394 |
| 2 | 26 | 42 | 380 | 82 | 855 | 122 | 1409 |
| 3 | 35 | 43 | 390 | 83 | 868 | 123 | 1424 |
| 4 | 45 | 44 | 401 | 84 | 881 | 124 | 1438 |
| 5 | 54 | 45 | 412 | 85 | 894 | 125 | 1453 |
| 6 | 63 | 46 | 423 | 86 | 907 | 126 | 1468 |
| 7 | 71 | 47 | 434 | 87 | 920 | 127 | 1483 |
| 8 | 79 | 48 | 445 | 88 | 934 | 128 | 1498 |
| 9 | 86 | 49 | 456 | 89 | 947 | 129 | 1513 |
| 10 | 93 | 50 | 467 | 90 | 960 | 130 | 1528 |
| 11 | 99 | 51 | 479 | 91 | 973 | 131 | 1543 |
| 12 | 106 | 52 | 490 | 92 | 987 | 132 | 1558 |
| 13 | 112 | 53 | 501 | 93 | 1000 | 133 | 1573 |
| 14 | 118 | 54 | 513 | 94 | 1014 | 134 | 1588 |
| 15 | 125 | 55 | 524 | 95 | 1027 | 135 | 1603 |
| 16 | 132 | 56 | 536 | 96 | 1041 | 136 | 1619 |
| 17 | 140 | 57 | 547 | 97 | 1055 | 137 | 1634 |
| 18 | 147 | 58 | 559 | 98 | 1068 | 138 | 1649 |
| 19 | 155 | 59 | 571 | 99 | 1082 | 139 | 1664 |
| 20 | 164 | 60 | 582 | 100 | 1096 | 140 | 1680 |
| 21 | 172 | 61 | 594 | 101 | 1109 | 141 | 1695 |
| 22 | 181 | 62 | 606 | 102 | 1123 | 142 | 1711 |
| 23 | 190 | 63 | 618 | 103 | 1137 | 143 | 1726 |
| 24 | 199 | 64 | 630 | 104 | 1151 | 144 | 1742 |
| 25 | 209 | 65 | 642 | 105 | 1165 | 145 | 1757 |
| 26 | 218 | 66 | 654 | 106 | 1179 | 146 | 1773 |
| 27 | 228 | 67 | 666 | 107 | 1193 | 147 | 1788 |
| 28 | 237 | 68 | 679 | 108 | 1207 | 148 | 1804 |
| 29 | 247 | 69 | 691 | 109 | 1221 | 149 | 1820 |
| 30 | 257 | 70 | 703 | 110 | 1236 | 150 | 1835 |
| 31 | 266 | 71 | 716 | 111 | 1250 | 151 | 1851 |
| 32 | 276 | 72 | 728 | 112 | 1264 | 152 | 1867 |
| 33 | 286 | 73 | 740 | 113 | 1278 | 153 | 1883 |
| 34 | 296 | 74 | 753 | 114 | 1293 | 154 | 1899 |
| 35 | 307 | 75 | 766 | 115 | 1307 | 155 | 1915 |
| 36 | 317 | 76 | 778 | 116 | 1322 | 156 | 1931 |
| 37 | 327 | 77 | 791 | 117 | 1336 | 157 | 1947 |
| 38 | 338 | 78 | 804 | 118 | 1351 | 158 | 1963 |
| 39 | 348 | 79 | 816 | 119 | 1365 | 159 | 1979 |
| 40 | 358 | 80 | 829 | 120 | 1380 | 160 | 1995 |

So we have

$$E[S] = k\mu$$

$$\text{Var}[S] = k\sigma^2 \tag{A.2}$$

$$E[S^2] = (E[S])^2 + \text{Var}[S] = k^2\mu^2 + k\sigma^2.$$

Then, from (6) we get

$$E[\widehat{N}_R] = \frac{E[S^2]}{Z_\alpha^2} - k = \frac{k^2\mu^2 + k\sigma^2}{Z_\alpha^2} - k$$

$$\text{Var}[\widehat{N}_R] = \frac{\text{Var}[S^2]}{Z_\alpha^4} = \frac{E[S^4] - (E[S^2])^2}{Z_\alpha^4}. \tag{A.3}$$

TABLE 5: Moments of the Normal distribution with mean $\mu$ and variance $\sigma^2$.

| Order | Noncentral moment | Central moment |
|---|---|---|
| 1 | $\mu$ | 0 |
| 2 | $\mu^2 + \sigma^2$ | $\sigma^2$ |
| 3 | $\mu^3 + 3\mu\sigma^2$ | 0 |
| 4 | $\mu^4 + 6\mu^2\sigma^2 + 3\sigma^4$ | $3\sigma^4$ |
| 5 | $\mu^5 + 10\mu^3\sigma^2 + 15\mu\sigma^4$ | 0 |

TABLE 6: Moments of the Poisson distribution with parameter $\lambda$.

| Order | Noncentral moment | Central moment |
|---|---|---|
| 1 | $\lambda$ | $\lambda$ |
| 2 | $\lambda + \lambda^2$ | $\lambda$ |
| 3 | $\lambda + 3\lambda^2 + \lambda^3$ | $\lambda$ |
| 4 | $\lambda + 7\lambda^2 + 6\lambda^3 + \lambda^4$ | $\lambda + 3\lambda^2$ |
| 5 | $\lambda + 15\lambda^2 + 25\lambda^3 + 10\lambda^4 + \lambda^5$ | $\lambda + 10\lambda^2$ |

Now we seek to compute $E[S^4]$, $E[S^2]$. For this we need the moments of the normal distribution, which are given in Table 5 [59].

So

$$
\begin{aligned}
&\text{Var}\left[\widehat{N}_R\right] \\
&= \frac{k^4\mu^4 + 6k^3\mu^2\sigma^2 + 3k^2\sigma^4 - \left(k^2\mu^2 + k\sigma^2\right)^2}{Z_\alpha^4} \\
&= \frac{4k^3\mu^2\sigma^2 + 2k^2\sigma^4}{Z_\alpha^4} \\
&\implies \text{Var}\left[\widehat{N}_R\right] = \frac{2k^2\sigma^2\left(2k\mu^2 + \sigma^2\right)}{Z_\alpha^4}.
\end{aligned}
\tag{A.4}
$$

*(b) Random k.* In this approach, we additionally assume that $k \sim \text{Pois}(\lambda)$. So $S$ is a compound Poisson distributed variable [60]. Hence, from the law of total expectation and the law of total variance [44], we get

$$
\begin{aligned}
E[S] &= E[k]\,E[Z_i] = \lambda\mu \\
\text{Var}[S] &= E[k]\,E[Z_i^2] = \lambda\left(\mu^2 + \sigma^2\right).
\end{aligned}
\tag{A.5}
$$

Thus, from (6) we get

$$
\begin{aligned}
E\left[\widehat{N}_R\right] &= \frac{E\left[S^2\right]}{Z_\alpha^2} - E[k] \\
&= \frac{(E[S])^2 + \text{Var}[S]}{Z_\alpha^2} - \lambda \\
&\implies E\left[\widehat{N}_R\right] = \frac{\lambda^2\mu^2 + \lambda\left(\mu^2 + \sigma^2\right)}{Z_\alpha^2} - \lambda, \\
\text{Var}\left[\widehat{N}_R\right] &= \frac{\text{Var}\left[S^2\right]}{Z_\alpha^4} + \text{Var}[k] - 2 \cdot \frac{\text{Cov}\left[S^2, k\right]}{Z_\alpha^2} \\
&= \frac{E\left[S^4\right] - \left(E\left[S^2\right]\right)^2}{Z_\alpha^4} \\
&\quad + \lambda - 2 \cdot \frac{E\left[kS^2\right] - E[k]\,E\left[S^2\right]}{Z_\alpha^2}.
\end{aligned}
\tag{A.6}
$$

To compute the final variance, it is more convenient to compute each component separately.

We will need the moments of a Poisson distribution [60], which are given in Table 6.

We then have

$$
\begin{aligned}
E\left[S^4\right] &= E\left[E\left[S^4 \mid k\right]\right] \\
&= E\left[k^4\mu^4 + 6k^3\mu^2\sigma^2 + 3k^2\sigma^4\right] \\
&= \mu^4 E\left[k^4\right] + 6\mu^2\sigma^2 E\left[k^3\right] + 3\sigma^4 E\left[k^2\right] \\
&\implies E\left[S^4\right] = \left(\lambda^4 + 6\lambda^3 + 7\lambda^2 + \lambda\right)\mu^4 \\
&\quad + 6\left(\lambda^3 + 3\lambda^2 + \lambda\right)\mu^2\sigma^2 + 3\left(\lambda^2 + \lambda\right)\sigma^4 \\
E\left[S^2\right] &= (E[S])^2 + \text{Var}[S] \\
&= \lambda^2\mu^2 + \lambda\left(\mu^2 + \sigma^2\right) \\
&= \left(\lambda^2 + \lambda\right)\mu^2 + \lambda\sigma^2 \\
\left(E\left[S^2\right]\right)^2 &= \left(\lambda^4 + 2\lambda^3 + \lambda^2\right)\mu^4 \\
&\quad + 2\left(\lambda^3 + \lambda^2\right)\mu^2\sigma^2 + \lambda^2\sigma^4.
\end{aligned}
\tag{A.7}
$$

So

$$
\begin{aligned}
&E\left[S^4\right] - \left(E\left[S^2\right]\right)^2 \\
&= \left(4\lambda^3 + 6\lambda^2 + \lambda\right)\mu^4 \\
&\quad + \left(4\lambda^3 + 16\lambda^2 + 6\lambda\right)\mu^2\sigma^2 + \left(2\lambda^2 + 3\lambda\right)\sigma^4.
\end{aligned}
\tag{A.8}
$$

Also

$$
\begin{aligned}
E\left[kS^2\right] &= E\left[E\left[kS^2 \mid k\right]\right] = E\left[kE\left[S^2 \mid k\right]\right] \\
&= E\left[k\left(k^2\mu^2 + k\sigma^2\right)\right] \\
&= \mu^2 E\left[k^3\right] + \sigma^2 E\left[k^2\right] \\
&= \left(\lambda^3 + 3\lambda^2 + \lambda\right)\mu^2 + \left(\lambda^2 + \lambda\right)\sigma^2 \\
E\left[kS^2\right] - E[k]\,E\left[S^2\right] &= \left(2\lambda^2 + \lambda\right)\mu^2 + \lambda\sigma^2.
\end{aligned}
\tag{A.9}
$$

Hence, we finally have

$$
\begin{aligned}
\operatorname{Var}\left[\widehat{N}_R\right] = &\left(\left(4\lambda^3 + 6\lambda^2 + \lambda\right)\mu^4\right.\\
&+ \left(4\lambda^3 + 16\lambda^2 + 6\lambda\right)\mu^2\sigma^2\\
&+ \left.\left(2\lambda^2 + 3\lambda\right)\sigma^4\right)\\
&\times \left(Z_\alpha^4\right)^{-1} - 2 \cdot \frac{\left(2\lambda^2 + \lambda\right)\mu^2 + \lambda\sigma^2}{Z_\alpha^2} + \lambda.
\end{aligned}
\tag{A.10}
$$

## B. Proof of Expression (14): The Characteristic Function

From (13) we have that

$$
\begin{aligned}
\psi_{N_R}(t) &= E\left[\exp\left(itN_R\right)\right]\\
&= \int_0^{+\infty} \exp\left(itn_R\right) f_{N_R}\left(n_R\right) dn_R\\
&= \int_0^{+\infty} \frac{Z_\alpha}{2\Phi\left(\lambda^*\right)\sqrt{2\pi k\sigma^2\left(n_R + k\right)}}\\
&\quad \times \exp\left[itn_R - \frac{\left(Z_\alpha\sqrt{n_R + k} - k\mu^2\right)}{2k\sigma^2}\right] dn_R\\
&\qquad\qquad \left(\text{let } w = Z_\alpha\sqrt{n_R + k}\right)\\
&= \frac{1}{\Phi\left(\lambda^*\right)} \int_{Z_\alpha\sqrt{k}}^{+\infty} \frac{1}{\sqrt{2\pi k\sigma^2}}\\
&\quad \times \exp\left[it\left(\frac{w^2}{Z_\alpha^2} - k\right) - \frac{\left(w - k\mu\right)^2}{2\sigma^2}\right] dw\\
&\qquad\qquad \left(\text{let } y = \frac{w - k\mu}{\sqrt{n}\sigma}\right)\\
&= \frac{1}{\Phi\left(\lambda^*\right)} \int_{-\lambda^*}^{+\infty} \frac{1}{\sqrt{2\pi}}\\
&\quad \times \exp\left[it\left(\frac{k\sigma^2 y^2 + 2n\sqrt{k}\mu\sigma y + k^2\mu^2}{Z_\alpha^2} - k\right)\right.\\
&\qquad\qquad \left. - \frac{y^2}{2}\right] dy\\
&= \frac{\exp\left(-kit\right)}{\Phi\left(\lambda^*\right)} \int_{-\lambda^*}^{+\infty} \frac{1}{\sqrt{2\pi}}\\
&\quad \times \exp\left[\frac{2k\sigma^2 it - Z_\alpha^2}{2Z_\alpha^2} y^2\right.\\
&\qquad\qquad \left. + \frac{2k\sqrt{k}\mu\sigma it}{Z_\alpha^2} y + \frac{k^2\mu^2 it}{Z_\alpha^2}\right] dy
\end{aligned}
$$

$$
\left(\text{let } \mu_1 = \frac{2k\sqrt{k}\mu\sigma it}{Z_\alpha^2 - 2k\sigma^2 it}, \sigma_1^2 = \frac{Z_\alpha^2}{Z_\alpha^2 - 2k\sigma^2 it}\right)
$$

$$
\begin{aligned}
&= \frac{\exp\left(-kit\right)}{\Phi\left(\lambda^*\right)} \int_{-\lambda^*}^{+\infty} \frac{1}{\sqrt{2\pi}}\\
&\quad \times \exp\left[-\frac{\left(y - \mu_1\right)^2}{2\sigma_1^2} + \frac{k^2\mu^2 it}{Z_\alpha^2 - 2k\sigma^2 it}\right] dy\\
&\qquad\qquad \left(\text{let } x = \frac{y - \mu_1}{\sigma_1}\right)\\
&= \frac{\sigma_1 \exp\left(k^2\mu^2 it/\left(Z_\alpha^2 - 2k\sigma^2 it\right) - kit\right)}{\Phi\left(\lambda^*\right)}\\
&\quad \times \int_{\left(-\mu_1 - \lambda^*\right)/\sigma_1}^{+\infty} \frac{1}{\sqrt{2\pi}} \exp\left(-\frac{x^2}{2}\right) dx\\
&= \frac{Z_\alpha \exp\left(k^2\mu^2 it/\left(Z_\alpha^2 - 2n\sigma^2 it\right) - kit\right)}{\Phi\left(\lambda^*\right)\left(Z_\alpha^2 - 2n\sigma^2 it\right)^{1/2}}\\
&\quad \times \left[\Phi\left(+\infty\right) - \Phi\left(\frac{-\mu_1 - \lambda^*}{\sigma_1}\right)\right]\\
\Longrightarrow \psi_{N_R}(t) &= \frac{\Phi\left(\left(\mu_1 + \lambda^*\right)/\sigma_1\right)}{\Phi\left(\lambda^*\right)}\\
&\quad \cdot \frac{Z_\alpha \exp\left(k^2\mu^2 it/\left(Z_\alpha^2 - 2\sigma^2 it\right) - kit\right)}{\left(Z_\alpha^2 - 2k\sigma^2 it\right)^{1/2}}
\end{aligned}
\tag{B.1}
$$

because $\Phi(+\infty) - \Phi((-\mu_1 - \lambda^*)/\sigma_1) = 1 - \Phi(-(\mu_1 + \lambda^*)/\sigma_1) = \Phi((\mu_1 + \lambda^*)/\sigma_1)$.

## C. Proof of Expressions (15) and (16)

The cumulant generating function is

$$
\begin{aligned}
g(t) &= \ln\left[\psi_{N_R}(-it)\right]\\
&= \frac{k^2\mu^2 t}{Z_\alpha^2 - 2k\sigma^2 t}\\
&\quad - kt - \frac{1}{2}\ln\left(Z_\alpha^2 - 2k\sigma^2 t\right)\\
&\quad + \ln\left[\Phi\left(\frac{\mu_1 + \lambda^*}{\sigma_1}\right)\right] + \ln\frac{Z_\alpha}{\Phi\left(\lambda^*\right)}, \quad t < \frac{Z_\alpha^2}{2k\sigma^2}.
\end{aligned}
\tag{C.1}
$$

For simplicity, let $\Delta = (\mu_1 + \lambda^*)/\sigma_1$. Then

$$
g'(t) = \frac{k^2\mu^2 Z_\alpha^2}{\left(Z_\alpha^2 - 2k\sigma^2 t\right)^2} - k + \frac{k\sigma^2}{Z_\alpha^2 - 2k\sigma^2 t} + \frac{\phi(\Delta)}{\Phi(\Delta)}\Delta',
\tag{C.2}
$$

with $\Delta' = 2k\sqrt{k}\mu\sigma Z_\alpha/(Z_\alpha^2 - 2k\sigma^2 t)^{3/2} - k\sigma^2 \lambda^*/Z_\alpha(Z_\alpha^2 - 2k\sigma^2 t)^{1/2}$.

Then, $g'(0)$ leads to (15).
Next

$$g''(t) = \frac{4k^3\mu^2\sigma^2 Z_\alpha^2}{\left(Z_\alpha^2 - 2k\sigma^2 t\right)^3} + \frac{2k^2\sigma^4}{\left(Z_\alpha^2 - 2k\sigma^2 t\right)^2}$$
$$\quad + \frac{\phi(\Delta)}{\Phi(\Delta)}\left[-\frac{\phi(\Delta)}{\Phi(\Delta)}\Delta'^2 - \Delta\Delta'^2 + \Delta''\right]. \quad \text{(C.3)}$$

Then $g''(0)$ leads to (16).

## Conflict of Interests

The authors declare that there is no conflict of interests regarding the publication of this paper.

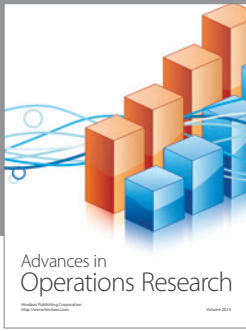

Advances in
Operations Research

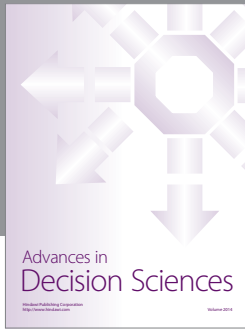

Advances in
Decision Sciences

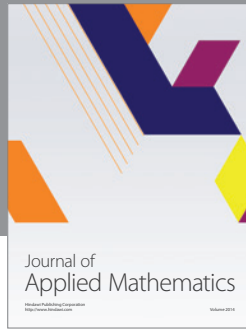

Journal of
Applied Mathematics

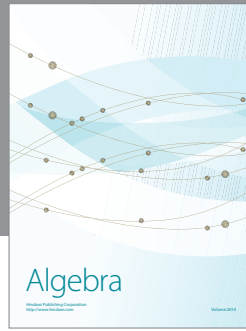

Algebra

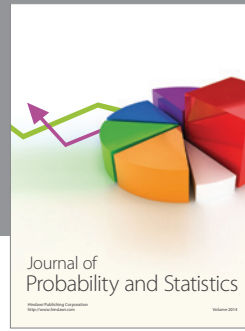

Journal of
Probability and Statistics

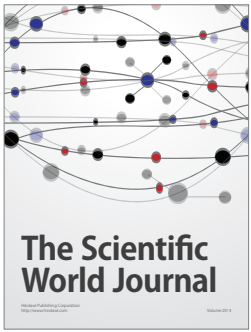

The Scientific
World Journal

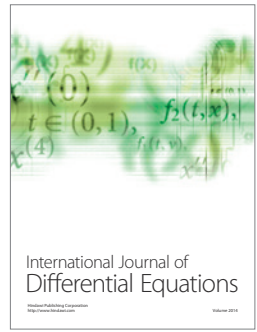

International Journal of
Differential Equations

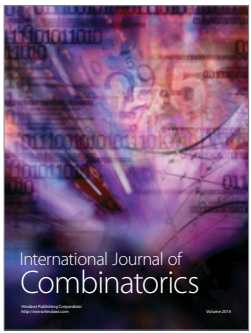

International Journal of
Combinatorics

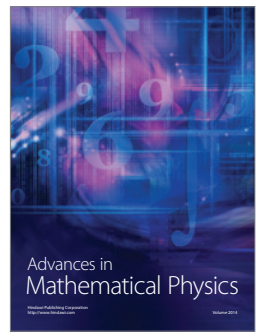

Advances in
Mathematical Physics

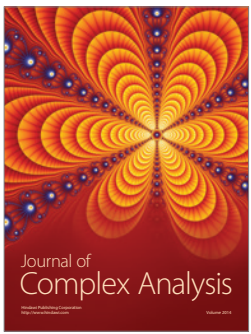

Journal of
Complex Analysis

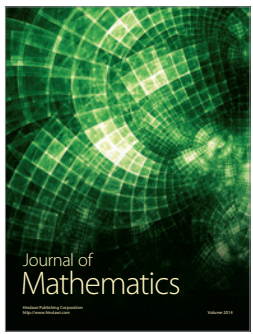

Journal of
Mathematics

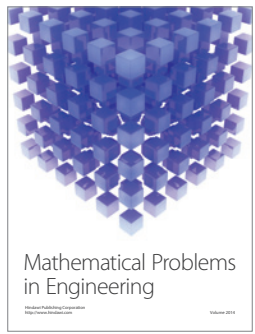

Mathematical Problems
in Engineering

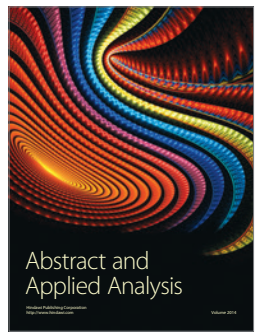

Abstract and
Applied Analysis

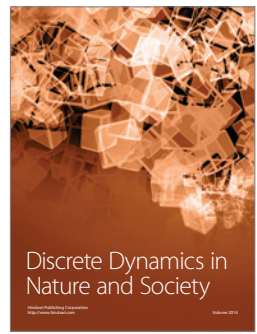

Discrete Dynamics in
Nature and Society

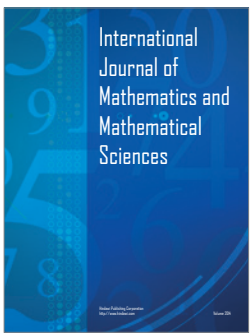

International
Journal of
Mathematics and
Mathematical
Sciences

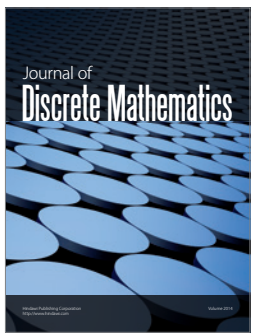

Journal of
Discrete Mathematics

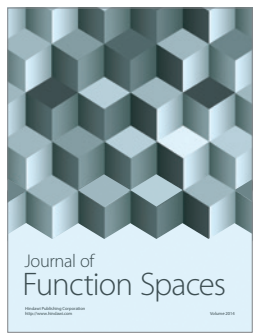

Journal of
Function Spaces

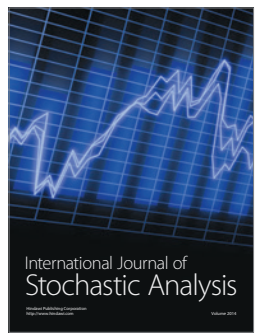

International Journal of
Stochastic Analysis

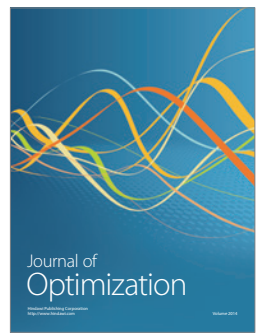

Journal of
Optimization

Submit your manuscripts at
http://www.hindawi.com

Hindawi